\documentclass{JHEP3}
\title{Noncommutative fields and actions\\ of twisted Poincar\'e
algebra}

\author{M. Chaichian$ ^a$, P. P. Kulish$ ^{a, b}$,
A. Tureanu$ ^a$, R. B. Zhang$ ^c$, Xiao Zhang$ ^d$\\
$ ^a$High Energy Physics Division, Department of Physical Sciences,
University of Helsinki and Helsinki Institute of Physics, P.O. Box
64, 00014 Helsinki, Finland;\\

$ ^b$St.Petersburg Department of Steklov Institute of
Mathematics, Fontanka, 27, St. Petersburg, 191023, Russia;\\

$ ^c$School of Mathematics and Statistics, University of Sydney, Sydney,
NSW 2006, Australia;\\

$ ^d$Institute of Mathematics, Academy of Mathematics and Systems
Science, Chinese Academy of Sciences, Beijing, China}

\abstract{Within the context of the twisted Poincar\'e algebra,
there exists no noncommutative analogue of the Minkowski space
interpreted as the homogeneous space of the Poincar\'e group
quotiented by the Lorentz group. The usual definition of commutative
classical fields as sections of associated vector bundles on the
homogeneous space does not generalise to the noncommutative setting,
and the twisted Poincar\'e algebra does not act on
noncommutative fields in a canonical way. We make a tentative proposal for the
definition of noncommutative classical fields of any spin over the Moyal
space, which has the desired representation theoretical properties.
We also suggest a way to search for noncommutative Minkowski spaces
suitable for studying noncommutative field theory with deformed
Poincar\'e symmetries.}


\begin{document}


\def\1{\hbox{1\kern-.35em\hbox{1}}}


\newtheorem{theorem}{Theorem}[section]
\newtheorem{lemma}{Lemma}[section]
\newtheorem{proposition}{Proposition}[section]
\newtheorem{corollary}{Corollary}[section]
\newtheorem{definition}{Definition}[section]
\newtheorem{remark}{Remark}[section]
\newtheorem{remarks}{Remarks}[section]

\newcommand{\st}[1]{\ensuremath{^{\scriptstyle \textrm{#1}}}}


\newcommand{\Z}{{\mathbb Z}}
\newcommand{\Q}{{\mathbb Q}}
\newcommand{\R}{{\mathbb R}}
\newcommand{\C}{{\mathbb C}}

\newcommand{\gl}{{{\mathfrak{gl}}_{m|n}}}
\newcommand{\Uq}{{{\rm U}_q}(\fg)}
\newcommand{\Ugl}{{{\rm U}_q}(\gl)}

\newcommand{\fb}{{\mathfrak b}}
\newcommand{\fg}{{\mathfrak g}}
\newcommand{\fh}{{\mathfrak h}}
\newcommand{\fj}{{\mathfrak j}}
\newcommand{\fk}{{\mathfrak k}}
\newcommand{\fl}{{\mathfrak l}}
\newcommand{\fn}{{\mathfrak n}}
\newcommand{\fp}{{\mathfrak p}}
\newcommand{\fq}{{\mathfrak q}}
\newcommand{\fr}{{\mathfrak r}}
\newcommand{\ft}{{\mathfrak t}}
\newcommand{\fu}{{\mathfrak u}}
\newcommand{\fw}{{\mathfrak w}}
\newcommand{\fx}{{\mathfrak x}}

\newcommand{\bw}{{\bf w}}
\newcommand{\bm}{{\bf m}}
\newcommand{\bk}{{\bf k}}
\newcommand{\ba}{{\bf a}}

\newcommand{\ad}{{\rm ad}}
\newcommand{\id}{{\rm id}}
\newcommand{\Hom}{{\rm Hom}}
\newcommand{\End}{{\rm End}}
\newcommand{\I}{{\rm I}}
\newcommand{\U}{{\rm U}}
\newcommand{\cA}{{\mathcal A}}
\newcommand{\cC}{{\mathcal C}}
\newcommand{\cE}{{\mathcal E}}
\newcommand{\cF}{{\mathcal F}}
\newcommand{\cH}{{\mathcal H}}
\newcommand{\cI}{{\mathcal I}}
\newcommand{\cS}{{\mathcal S}}
\newcommand{\cU}{{\mathcal U}}
\newcommand{\cO}{{\mathcal O}}

\newcommand{\CR}{{C^\infty(\R^{1, 3})}}

\newcommand{\hR}{{\hat R}}
\newcommand{\hL}{{\hat L}}
\newcommand{\hDelta}{{\hat\Delta}}

\newcommand{\tL}{{\tilde L}}
\newcommand{\tR}{{\tilde R}}

\newcommand{\e}{{\epsilon}}

\newcommand{\cR}{{\mathcal R}}
\newcommand{\cL}{{\mathcal L}}


\section{Introduction}\label{introduction}

There have been intensive research activities in quantum field
theory on noncommutative spaces (see, e.g., \cite{SWit, DN} and
references therein) in recent years. All aspects of
noncommutative quantum field theory on the Moyal space have been
studied, which include foundational issues, renormalisation as well
as model building for particle physics.
We mention in particular that noncommutative quantum field theories
behave very differently from their commutative counterparts,
as can be seen, e.g., from the UV/IR mixing \cite{MRS} appearing in
the noncommutative case.

A major conceptual advance was the recognition \cite{CKNT} that the
twisted Poincar\'e algebra should play the same role in
noncommutative quantum field theory on the Moyal space as that
played by the Poincar\'e group in usual relativistic quantum field
theory. The merit of the twisted Poincar\'e symmetry of the
noncommutative QFT is that its particle representations are
identical with the ones of the usual Poincar\'e symmetry, since the
structure of the twisted Poincar\'e algebra is identical to the one
of the Poincar\'e algebra  and hence the Casimir operators are the
same. As a result, the particle states of NC QFT are classified
according to their mass and spin \cite{CKNT} as ordinarily. The
study of the consequences of this twisted Poincar\'e symmetry
\footnote{Recently several papers \cite{FW, JM, BPQ2} claimed that
twisted Poincar\'e invariant noncommutative quantum field theory on
the Moyal space had the same $S$-matrix as its commutative
counterpart. This is very surprising in view of the drastic
differences between the commutative and noncommutative theories.}
has increased the interest in the subject since the publication of
\cite{CKNT}. Attempts \cite{ADMW}  have also been made to gauge the
twisted Poincar\'e algebra in order to construct a noncommutative
theory of general relativity. Other possible noncommutative
spacetime symmetries have also been studied in the literature, e.g.
the $\kappa$-Poincar\'e algebra \cite{LNR}.

The twisted Poincar\'e invariance of noncommutative quantum
field theory is an extremely important issue, which should be investigated
systematically by starting from first principles. To consider it,
one needs to have a representation theoretical
interpretation of the fields and also a precise definition of the
actions of the twisted Poincar\'e algebra on them. Unfortunately
neither is well understood, especially for fields with nonzero spin (of
course we can not give precise meanings to the terms ``fields" and
``spin" yet). In the literature there is enough material for one to
extract a general definition of a classical noncommutative scalar field on
the Moyal space and specify the precise transformation rule for it under
the twisted Poincar\'e algebra (see, e.g., \cite{{CPT}} and also
later treatments by other authors). However, there is hardly any
discussion on what fields with nonzero spin should be, leave alone
any precise formulation, from first principles, of their
transformation rules under the twisted Poincar\'e algebra. Some
researchers are aware of aspects of this problem. For example, there
was a lengthy discussion in \cite{FW} on the need of formulating a
transformation rule of fields under the twisted Poincar\'e algebra,
but the authors did not address the issue directly, rather they
suggested a way to side-step it instead. Also the paper \cite{D}
aimed at addressing similar issues for general Hopf algebras.

For simplicity we shall consider twisted Poincar\'e invariance of
noncommutative classical field theory. Recall that in the commutative
setting, Minkowski space is realised as the quotient of the
Poincar\'e group by the Lorentz group, and a classical field is a
section of a vector bundle induced by some representation of the
Lorentz group (or its double cover). The space of sections of the
bundle, which is the well-known induced module,
carries a natural action of the entire Poincar\'e group.
One would expect that the Moyal space and noncommutative
fields on it should be understood in such terms as well.

We shall carefully examine the induced module construction in
Section \ref{classical}, and then investigate the possibility of
generalising it to the twisted Poincar\'e algebra in Section
\ref{noncommutative}. Unfortunately, we find that the natural
generalisation does not go through, primarily because the universal
enveloping algebra of the Lorentz Lie algebra is not a Hopf
subalgebra of the twisted Poincar\'e algebra. We shall explain in
detail the obstacle preventing the generalisation in the second half
of Section \ref{noncommutative}.  To further illustrate the
problems, we examine in Section \ref{Moyal} the two noncommutative
algebras in the literature which are closely related to the Moyal space
and arise from the representation theory of the twisted Poincar\'e
algebra, and explain why they are not useful for defining classical
fields. These rather unexpected difficulties indicate that one can not use the same canonical definitions in the case of noncommutative fields on the Moyal space to address the representation theoretical properties relative to the twisted Poincar\'e
algebra.

One could, however, approach the problem differently. In Section \ref{fields}, we propose a
definition of noncommutative classical fields which agrees with what
noncommutative scalar fields were implicitly taken to be in the literature
(see, e.g., \cite{CPT}), and recovers the usual definition of scalar
fields in the commutative case. We hope that the proposal will
provide a useful framework for studying twisted Poincar\'e
invariance of quantum field theory on the Moyal space.

A further useful aspect of results in this paper is that they
provide a theoretical basis for the search of noncommutative
analogues of the Minkowski space which are suitable for studying
noncommutative field theory with deformed Poincar\'e symmetries. We
shall discuss this point in more detail in Section \ref{conclusion}.

Before closing this section, we mention that we shall limit
ourselves to the case where the noncommutative fields carry no
internal degrees of freedom. This enables us to better focus on
properties of noncommutative fields relative to the twisted
Poincar\'e algebra. All results of this paper can be generalised to
include internal degrees of freedom in a straightforward manner.

\section{Induced modules of Poincar\'e group
and classical fields}\label{classical}

We review the basic definition of Poincar\'e group actions on
commutative classical fields. This relatively well-known material is
needed later when we investigate the possibilities/difficulties of
generalising it to the twisted Poincar\'e algebra.

Choose the metric $\eta=diag(-1, 1, 1, 1)$ for $\R^{1, 3}$. Let $G$
denote the Poincar\'e group, which is the semi-direct product of the
Lorentz group and the abelian group of translations on $\R^{1, 3}$,
where the Lorentz group is defined with respect to the metric
$\eta$. Since spinor fields should be included in the framework as
well, we consider instead the covering group
$\tilde G=Spin(1, 3)\ltimes \R^{1, 3}$, the
semi-direct product of $Spin(1, 3)$ with the group of translations.
Now $Spin(1, 3)$ acts on translations via the surjection $\pi$ from
$Spin(1, 3)$ to the Lorentz group. For convenience we shall denote
$Spin(1, 3)$ by $L$.

Denote the coordinate of $\R^{1, 3}$ by $x=(x^0, x^1, x^2, x^3)$.
For computational purposes it is the best to write an element of
$\tilde G$ as $\Lambda\exp(iPx)$ where $\Lambda\in L$ and
$x\in\R^{1, 3}$, with the product of two elements $\Lambda\exp(iPx)$
and $\Lambda'\exp(iPy)$ given by
\[
\Lambda'\exp(iPy) \Lambda\exp(iPx) = \Lambda'\Lambda \exp(iP
(\Lambda^{-1}(y) +x)).
\]
Here $\Lambda^{-1}(y)^\mu=\pi(\Lambda^{-1})_\nu^\mu y^\nu$.

Let $C^\infty(\tilde G)$ be the {\em set} of smooth functions on
$\tilde G$. In the present (untwisted) case, it forms a commutative
algebra under the usual pointwise multiplication from calculus.

\begin{remark}
The pointwise multiplication of functions on $\tilde G$ is
intimately related to the fact that we give the group algebra the
co-commutative co-multiplication
 \[\Delta_0(g)=g\otimes g\] for all elements
$g$ in the group. This co-multiplication is compatible with the
standard co-commutative co-multiplication (\ref{untwisted-comult})
for the Poincar\'e algebra.
\end{remark}

There are two natural actions of $\tilde G$ on $C^\infty(\tilde G)$,
the left and right translations, which we shall denote by $\cL$ and
$\cR$ respectively. For any $\phi\in C^\infty(\tilde G)$ and
$g\in\tilde G$, $\cL_g(\phi)$ and $\cR_g(\phi)$ are respectively
defined by
\[
\cL_g(\phi)(g_1) = \phi(g^{-1} g_1), \quad \cR_g(\phi)(g_1)=\phi(g_1
g), \quad \forall g_1\in\tilde G.
\]
It is these actions which give rise to actions of the Poincar\'e
group on classical fields. We shall carefully examine this point now
with the view of possible generalisations to the noncommutative
case.

To discuss properties of classical fields on $\R^{1, 3}$ in relation
to the Poincar\'e group, we first note that $\R^{1, 3}\cong \tilde
G/L$. At this point, we need to make a choice in interpreting this
either as a left or right coset space. We shall take $\tilde G/L$ as
the right coset space consisting of equivalence classes with the
following equivalence relation $\Lambda\exp(iPx) \sim \Lambda'
\exp(i Px)$ for all $\Lambda, \Lambda'\in L$.

Let $V$ be a finite dimensional $L$-module, and denote by $\rho$ the
representation of $L$ on $V$ relative to some choice of basis. Then
a classical field of a type characterised by $V$ is a section of the
associated $C^\infty$ vector bundle
$
\tilde G\times_L V \longrightarrow\tilde G/L.
$
Denote by $\Gamma(V)$ the space of the smooth sections of this
vector bundle, which is a subspace of $C^\infty(\tilde G)\otimes_\C
V$, where the latter vector space is endowed with an action of $L$
defined for any $\phi\otimes v\in C^\infty(\tilde G)\otimes_\C V$ by
\[
\Lambda(\phi\otimes v) = (\cL\otimes \rho)\Delta_0(\Lambda)(\phi) =
\cL_\Lambda(\phi)\otimes \rho(\Lambda)v, \quad \forall \Lambda\in L.
\]
Then $\Gamma(V)$ is the subspace of invariants of $C^\infty(\tilde
G)\otimes_\C V$ with respect to this $L$-action, that is,
\begin{eqnarray}\label{Gamma}
\Gamma(V) = \left(C^\infty(\tilde G)\otimes_\C V\right)^L.
\end{eqnarray}
As is well known, the space $\Gamma(V)$ of sections forms a module,
the induced module,  for the entire
Poincar\'e group $\tilde G$ defined for any $\Phi\in\Gamma(V)$ by
\begin{eqnarray}\label{action}
g(\Phi) = (\cR_g\otimes\id_V)\Phi, \quad g\in\tilde G.
\end{eqnarray}

\begin{remark}
Note that in formulating (\ref{Gamma}), it is of crucial importance
that the group algebra of $L$ is a Hopf subalgebra of the group
algebra of $\tilde G$ under the co-multiplication $\Delta_0$.
\end{remark}

Both equations (\ref{Gamma}) and (\ref{action}) can be made more
explicit. If $\Phi\in \Gamma(V)$, then (\ref{Gamma}) implies that
\begin{eqnarray} \label{invariance}
(\cL_\Lambda\otimes\id_V)\Phi = (\id\otimes \rho(\Lambda^{-1}))\Phi,
\quad \forall \Lambda\in L,
\end{eqnarray}
where the $\id$ on the right side is the identity map on
$C^\infty(\tilde G)$.  Therefore,
\[
\Phi(\Lambda\exp(iPx)) = \rho(\Lambda) \Phi(\exp(iPx)).
\]
Denote
\begin{eqnarray*}
\phi(x)=\Phi(\exp(iP x)), \quad (g\cdot\phi)(x)=
(g(\Phi))(\exp(iPx)), \quad g\in\tilde G.
\end{eqnarray*}
Then by using (\ref{invariance}), we easily see (with notations as
above) that
equation (\ref{action}) is equivalent to
\begin{eqnarray} \label{rule}\label{classical-action}
(\Lambda\exp(i P a)\cdot \phi)(x) = \rho(\Lambda) \phi(\Lambda^{-1}
x + a), \quad \Lambda\exp(i P a)\in\tilde G.
\end{eqnarray}

This is the familiar transformation rule for a classical field on
$\R^{1, 3}$ under the action of the Poincar\'e group. The type of a
classical field is determined by $V$. For example, the field is a
vector if $V$ is the natural module for the Lorentz group, and a
spinor if $V$ is a spinor module.

\begin{remark}
Quantum fields obey the quantum version of the transformation rule
(\ref{rule}) as one can see from equations (5.1.6) and (5.1.7) in
\cite{W}. Recall that in \cite{W}, relativistic quantum fields are
constructed through cluster decompositions of multi-particle states,
thus are of a secondary nature. Their transformation rule under the
Poincar\'e group is derived from the invariance of the scattering
matrix.
\end{remark}

Let $V$ and $V'$ be $L$-modules. If $f: V\longrightarrow V'$ is an
$L$-module homomorphism, it induces a morphism between the
associated vector bundles
\[ f_*: \Gamma(V)\longrightarrow \Gamma(V'), \quad
\Phi\mapsto (\id\otimes f)\Phi.
\]
Also note that the multiplication of $\C^\infty(\tilde G)$ induces a
tensor product map for the associated vector bundles
\begin{eqnarray}\label{tensor}
\Gamma(V)\otimes \Gamma(V') \longrightarrow \Gamma(V\otimes V'),
\end{eqnarray}
defined for $\Phi=\sum_i\phi_i\otimes v_i\in\Gamma(V)$ and
$\Psi=\sum_j\psi_j\otimes v'_j\in\Gamma(V')$ by
\[\Phi\otimes\Psi \mapsto \Phi\Psi =
\sum_{i, j} \phi_i\psi_j\otimes v_i\otimes v'_j.\]
By a direct computation one can show that the right hand side indeed
belongs to $\Gamma(V\otimes V')$. There is an obvious generalisation
of the map to more than two bundles.

Given a classical field $\Phi\in\Gamma(V)$, we may consider, say,
$\Phi^{k}\in\Gamma(V^{\otimes k})$. If there exists a module map $f$
from $V^{\otimes k}$ to the $1$-dimensional trivial $L$-module $\C$,
then $f_*(\Phi^k)$ is a complex valued function on $\R^{1, 3}$.
Then for all $\Lambda\exp(iPa)\in
\tilde G$,
\begin{eqnarray*}
\int dx (\Lambda\exp(iPa)\cdot (f_*(\Phi^k)))(x) &=& \int dx
(f_*(\Phi^k))(\Lambda^{-1}x+a)\\
&=&\int dx (f_*(\Phi^k))(x),
\end{eqnarray*}
that is, the integral $\int dx (f_*(\Phi^k))(x)$ (which means $\int
dx (f_*(\Phi^k))(\exp(iPx))$)  is Poincar\'e invariant. The
construction of the invariant integral can obviously generalise to
the case with more than one classical field which can be sections
of different vector bundles on $\tilde G/L$ (derivatives of a
section are considered as a sections of different vector bundles).
This is how one constructs Poincar\'e invariant Lagrangians in
classical field theory.

\begin{remark}
Unitarity of the induced module $\Gamma(V)$ is required in order to
have a sensible field theory.
\end{remark}

\section{Induced modules for the twisted Poincar\'e
algebra}

\subsection{Generalities on induced modules for the twisted Poincar\'e
algebra}\label{noncommutative}

In this section we shall first discuss induced modules of the
twisted Poincar\'e algebra in general terms, then explain the
obstruction preventing the generalisation of the constructions of
Section \ref{classical} to the noncommutative setting. The general
method of this section is adapted from the paper \cite{GZ} on a
geometric representation theory for quantum groups.

Let $\fg$ be the complexification of the Lie algebra $Lie(\tilde G)$
of the Poincar\'e group. Then $\fg=\fl+\fp$, where $\fl$ is the
complexification of the Lie algebra of the Lorentz group, and $\fp$
is the complexification of the Lie algebra of the group of
translations on $\R^{1, 3}$. A basis for $\fg$ is $\{J_{\mu \nu}, \
P_\mu \mid \mu, \nu =0, 1, 2, 3\}$ with the following commutation
relations
\begin{eqnarray}
\begin{array}{r l l}
{[J_{\mu \nu}, J_{\sigma \rho}]}&=\frac{1}{i} (\eta_{\nu \sigma }
J_{\mu \rho} -\eta_{\mu \sigma} J_{\nu \rho}
- \eta_{\nu \rho} J_{\mu \sigma} + \eta_{\mu \rho} J_{\nu \sigma}  ), \\
{[J_{\mu \nu}, P_\sigma]}&= i\eta_{\mu \sigma} P_{\nu} - i \eta_{\nu
\sigma} P_{\mu},\\
{[P_\mu, P_\nu]}&=0.
\end{array}
\end{eqnarray}
We denote  by $\cU$ the universal enveloping algebra of the
Poincar\'e algebra $\fg$.  The standard co-commutative
co-multiplication $\Delta_0$ is given by
\begin{eqnarray}\label{untwisted-comult}
\Delta_0(X)=X\otimes 1 + 1\otimes X, \quad \forall X\in\fg.
\end{eqnarray}

The twisted Poincar\'e algebra is the associative algebra $\cU$
equipped with a twisted co-multiplication defined in the following
way. Let $\theta=\left(\theta^{\mu\nu}\right)$ be a real $4\times 4$
skew symmetric matrix. Set
\[ \cF= \exp(\sum_{\mu, \nu}\frac{1}{2}i\theta^{\mu\nu} P_\mu\otimes P_\nu), \]
which is understood as belonging to some appropriate completion of
$\cU\otimes\cU$. The twisted co-multiplication is then defined by
\begin{eqnarray}
\Delta: \cU \longrightarrow \cU\otimes\cU, \quad u\mapsto
\cF\Delta_0(u)\cF^{-1},
\end{eqnarray}
which is indeed co-associative as can be easily shown.
Now for any $\omega\in\fl$ and $P\in \fp$,
\begin{eqnarray}\label{Delta}
\begin{array}{r l}
\Delta(\omega)&=\omega\otimes 1 + 1\otimes\omega -\frac{1}{2}\sum
i\theta^{\mu
\nu}([\omega, P_\mu]\otimes P_\nu + P_\mu\otimes[\omega, P_\nu]), \\
\Delta(P)&=P\otimes 1 + 1 \otimes P.
\end{array}
\end{eqnarray}
If we also define the co-unit $\epsilon$ and antipode $S$ respectively by
$\epsilon(1)=1$, $\epsilon(X)=0$, and $S(X)=-X$  for all $X\in\fg$,
then $\cU$ is a Hopf algebra with co-multiplication $\Delta$.

It is worth mentioning that the $\cF$ used to twist $\Delta_0$ to
obtain the new co-multiplication $\Delta$ is an example of a special
type of gauge transformations in the powerful theory of quasi-Hopf
algebras \cite{Dr1, Dr2} (see also \cite{CP}). We refer to
\cite{CD1} for more details on twisting co-multiplications.

The co-algebra structure of $\cU$ induces a natural associative
algebra structure on the dual space $\cU^*$ of $\cU$. Since $\Delta$
is clearly noncocommutative, $\cU^*$ is noncommutative. Note that
$\cU^*$ is a huge object, which contains $C^\infty(\tilde G)$ as a
subspace in some appropriate sense. There exist two left actions
$ \cL, \cR: \cU\otimes\cU^* \longrightarrow \cU^* $
respectively defined for any $f\in\cU^*$ and $u\in\cU$ by
\begin{eqnarray}
\cL_u(f)(w)= f(S^{-1}(u)w), \quad \cR_u(f)(w)=f(w u), \quad \forall
w\in \cU.
\end{eqnarray}

Let $\cA(\fg)$ be either $\cU^*$ itself or an appropriate subalgebra
of it. In the latter case we require that for any nonzero $u\in\cU$,
there exists some $a\in\cA(\fg)$ such that $a(u)\ne 0$. Also,
$\cA(\fg)$ should be stable under both the left and right
translations, that is, $\cL_u(\cA(\fg)),
\cR_u(\cA(\fg))\subset\cA(\fg)$ for all $u\in\cU$. The algebra
$\cA(\fg)$ will be taken as defining some noncommutative space
following the general philosophy of noncommutative geometry
\cite{Co}.

Let $\cC$ be a two-sided co-ideal of $\cU$ satisfying $c(1)=0$
for all $c\in\cC$, where $1$ is the identity element of $\cU$. Being a two-sided co-ideal means
that $\Delta(\cC)\subset \cC\otimes \cU + \cU\otimes\cC$. Now define
\begin{eqnarray}
\cA(\fg, \cC) := \left\{f\in \cA(\fg) \mid \cL_c(f) =0, \ \forall
c\in\cC\right\}.
\end{eqnarray}
Then $\cA(\fg, \cC)$ is a subalgebra of $\cA(\fg)$. The proof of this is
quite illuminating. If $f, g\in\cA(\fg, \cC)$, then for all
$c\in\cC$ and $u\in\cU$, we have
\begin{eqnarray*}
\cL_c(fg)(u) &=& (f\otimes
g)\Delta(S^{-1}(c)u) \\ &=& \sum_{(c), (u)} \cL_{c_{(2)}}(f)(u_{(1)}) \cL_{c_{(1)}}(g)(u_{(2)}),
\end{eqnarray*}
where we have used Sweedler's notation \cite{Sw} for the co-multiplications of $c$
and $u$. Since $\cC$ is a two-sided co-ideal, we have
\[\cL_c(fg)(u) = 0, \quad \forall u\in\cU.
\]
The algebra $\cA(\fg, \cC)$ is taken as defining a noncommutative
analogue of some homogeneous space of $\tilde G$.

As far as we are aware, this is the definition of noncommutative
homogeneous spaces that requires the weakest conditions on $\cC$. If
we also want to develop a theory of induced representations similar
to that in the setting of Lie groups, we need to impose the stronger
condition that $\cC$ generates a Hopf subalgebra of $\cU$.

Now we make the assumption that $\cC$ generates a Hopf subalgebra
$\cH$ of $\cU$.  Then
\[
\cA(\fg, \cC) = \cA(\fg)^{\cL_\cH},
\]
which is the subalgebra of $\cA(\fg)$ consisting of the $\cH$ invariant elements.
Let $V$ be a finite dimensional $\cH$-module. We define the vector
space
\begin{eqnarray}\label{qGamma}
\Gamma(V) := \left\{ \zeta\in \cA(\fg)\otimes_\C V \left|
\sum_{(u)}(\cL_{u_{(1)}}\otimes u_{(2)})\zeta = \epsilon(u) \zeta,
\forall u\in\cH \right. \right\},
\end{eqnarray}
where we have used Sweedler's notation
$\Delta(u)=\sum_{(u)}u_{(1)}\otimes u_{(2)}$ for the
co-multiplication of $u$. Then $\Gamma(V)$ is a two-sided $\cA(\fg,
\cC)$-module under the multiplication in $\cA(\fg)$: for any
$a\in\cA(\fg, \fl)$ and $\zeta=\sum\phi_i\otimes v_i\in\Gamma(V)$,
both
\[
a\zeta = \sum a\phi_i\otimes v_i \quad \mbox{and} \quad  \zeta a =
\sum \phi_i a\otimes v_i
\]
belong to $\Gamma(V)$.

\begin{remark}
Both the definition of $\Gamma(V)$ and its $\cA(\fg, \fl)$-module
structures rely in a crucial way on the Hopf algebra structure of
$\cH$.
\end{remark}

It is important to observe that $\Gamma(V)$ forms a left
$\cU$-module under the action
\begin{eqnarray}
\cU\otimes \Gamma(V) \longrightarrow \Gamma(V), \quad
u\otimes\zeta\mapsto (\cR_u\otimes \id_V)\zeta.
\end{eqnarray}
Furthermore, if $V$ and $V'$ are both $\cH$-modules, then there
exists a map
\begin{eqnarray}\label{qtensor}
\Gamma(V)\otimes \Gamma(V')\longrightarrow \Gamma(V\otimes V')
\end{eqnarray}
defined in exactly the same way as (\ref{tensor}).

\begin{remark}
By imposing appropriate conditions on the algebra $\cA(\fg)$ we can
reproduce the results in Section \ref{classical} this way by using
the usual co-multiplication $\Delta_0$ for $\cU$.
\end{remark}

In order for the twisted Poincar\'e algebra to play a similar role
in noncommutative field theory as that played by the Poincar\'e
group in commutative field theory, it appears to be quite necessary
to have a noncommutative analogue of the construction of induced
representations given in Section \ref{classical}. It is that
construction which provides the definition of classical fields on
$\R^{1, 3}$ and also specifies the action of the Poincar\'e group on
them.

If we wish to generalise Section \ref{classical} to the
noncommutative setting, we have to take such a subspace $\cC$ of
$\cU$ that contains $\fl$ but not any nontrivial subspace of $\fp$.
However, in this case $\cC$ can {\em not} be a two-sided co-ideal as
one can easily see by inspecting the co-multiplication
(\ref{Delta}). For example, the natural choice $\cC=\fl$
does not give us a two-sided co-ideal. It
still makes sense to define $\cA(\fg, \fl)=\cA(\fg, \cC)$ in this case,
however, $\cA(\fg, \fl)$ will not be a subalgebra of
$\cU^*$ since the universal enveloping of $\fl$ is not a Hopf subalgebra
of $\cU$.

This means that conceptually we can not regard $\cA(\fg, \fl)$ as
defining a noncommutative geometry. An immediate practical problem
caused by this is the following.  If we follow the type of thinking
in Section \ref{classical}, we would like to interpret elements of
$\cA(\fg, \fl)$ as a ``scalar field". Since $\cA(\fg, \fl)$ is not
an algebra, we do not know how to multiply two ``fields" (or a
``field" with itself) together.

Now we consider the induced module construction. If $V$ is merely an
$\fl$-module, then the corresponding $\Gamma(V)$ as that in
(\ref{qGamma}) can not be defined. One way out is to take $V$ to be
a $\cU$-module with trivial $\fp$ action. Then at least we can
define a $\Gamma(V)$ by (\ref{qGamma}). Now the map (\ref{qtensor})
is not defined. Therefore we can not simply generalise the classical construction to build Lagrangians
from elements of $\Gamma(V)$
and defining Wightman functions in the corresponding quantum theory.

To summarise,

\begin{quote}
there does not exist a noncommutative analogue of $\R^{1, 3}=\tilde
G/L$ in terms of the twisted Poincar\'e algebra, and the induced
module construction for the Poincar\'e group in Section
\ref{classical} can not be generalised to the twisted setting.
\end{quote}

Therefore, one does not have a straightforward generalisation of the
definition of classical fields to the noncommutative setting.
Note that a similar situation is encountered in the case of the $\kappa$-Poincar\'e algebra  \cite{LNR}, for the same reason
that the enveloping algebra of the Lorentz subalgebra is
not a Hopf subalgebra.

\subsection{Representation theoretical constructs related to Moyal
space}\label{Moyal}

There are two noncommutative algebras in the literature which arise
from the representation theory of the twisted Poincar\'e algebra and
are related to the Moyal space. We discuss difficulties which one
encounters when trying to take any of these algebras as the algebra
of functions on some noncommutative space and develops field theory
on it. There are also various inaccurate statements concerning the
relationship between these algebras and the Moyal space in the
literature, which we hope to clarify here.
We should mention that it is not hard to deduce the material below
from appropriate mathematical sources, e.g., \cite{Sw}.

\subsubsection{A module algebra}\label{modalg} Consider an
indecomposable module $V=X\oplus \C \1$ for the Poincar\'e algebra
$\cU$, where $\C\1$ is a 1-dimensional submodule, and
$X=\oplus_{\mu=0}^3 \C x^\mu$ forms the natural module for $\fl$.
Explicitly, the $\cU$ action on $V$ is given by
\[
J_{\mu \nu} (x^\sigma) = \frac{1}{i}\left(\delta_\nu^\sigma x_\mu -
\delta_\mu^\sigma x_\nu \right), \quad P_\mu (x^\sigma) =
\frac{1}{i}\delta_{\mu}^\sigma \1, \quad Y(\1) =0,  \  \forall
Y\in\fg.
\]
Let $T(V)$ be the tensor algebra of $V$. Then
$T(V)=\sum_{k=0}^\infty T(V)_k$ with $T(V)_k=V^{\otimes k}$ and
$T(V)_0=\C$. Now $T(V)$ has a natural $\cU$-module structure with
respect to the {\em twisted} co-multiplication $\Delta$.

Let $\omega =\frac{1}{2}\sum\omega^{\mu \nu} J_{\mu \nu}$ and
$P=\sum c^\mu P_\mu$,  where $\omega^{\mu \nu}$ and $c^\mu$ are
complex numbers. Set $\omega^\mu_\nu = \sum_\sigma
\omega^{\mu\sigma}\eta_{\sigma\nu}$. For the following elements of
$V\otimes V$,
\begin{eqnarray*}
A^{\mu \nu} &:=& x^\mu\otimes x^\nu - x^\nu \otimes x^\mu -
i\theta^{\mu \nu}\1\otimes\1,\\
V^\mu &:=& x^\mu\otimes\1 - \1\otimes x^\mu,
\end{eqnarray*}
we have
\[
\begin{array}{r l c l}
&\Delta(\omega)A^{\mu \nu} = i\sum_\sigma(\omega^\mu_\sigma
A^{\sigma \nu} - i\omega^\nu_\sigma A^{\sigma \mu}), &\quad&
\Delta(P)A^{\mu \nu} =-
i \left(c^\mu V^\nu - c^\nu V^\mu\right),\\
&\Delta(\omega)V^\mu = i \sum_\sigma\omega^\mu_\sigma V^\sigma,
&\quad&  \Delta(P)V^\mu = 0.
\end{array}
\]
Also observe that the element $1 - \1\in T(V)_0\oplus T(V)_1$ is an
invariant. Therefore, the two-sided ideal $\cI$ of $T(V)$ generated
by all $A^{\mu \nu}$, $V^\mu$ and $1-\1$ is a $\cU$-submodule with
respect to the twisted co-multiplication. Define the unital
associative algebra
\begin{eqnarray}
\cA: = T(V)/\cI,
\end{eqnarray}
which admits a natural action of the twisted Poincar\'e algebra
$\cU$. The algebra $\cA$ frequently appears in the literature. It
may be regarded as generated by $x^\mu$ ($\mu=0, 1, 2, 3$) and the
identity subject to the relations
\begin{equation}\label{modulealgebra}
x^\mu x^\nu - x^\nu x^\mu = i\theta^{\mu\nu}.
\end{equation}

These are the same as the familiar relations satisfied by the
coordinate functions of the Moyal space. However, one can not simply
assign numerical values to $x^\mu$ to obtain numbers from elements
of $\cA$. This fact prevents one from constructing field theory by
using the algebra $\cA$ directly. If one enlarges \cite{FW} $\cA$ by
allowing for appropriate infinite sums, then the resulting algebra
may be isomorphic to the algebra of the Moyal space with $x^\mu$ mapped
to the coordinate functions. Then one may regard a field theory on
the Moyal space as defined on $\cA$ via this isomorphism.

\medskip

Let us now investigate the algebra $\cA$ a little further.  Since
the ideal $\cI$ is not homogeneous as the generators $A^{\mu \nu}$
are not, the $\Z_+$ grading of $T(V)$ does not descend to $\cA$, but
induces a filtration
\[
\cA_0\subset \cA_1\subset \cA_2\subset \dots,
\]
where $\cA_i=T(V)_{\le i} /  (\cI\cap T(V)_{\le i})$ and $T(V)_{\le
i}=\sum_{k\le i}T(V)_k$. Every $\cA_i$ is obviously a
$\cU$-submodule, thus \[gr\cA_i:=\cA_i/\cA_{i-1}\] admits a natural
$\cU$-action. Then $gr\cA=\sum_i gr\cA_i$ is a graded algebra with
the momentum operators $P_\mu$ acting on it by zero, and the Lorentz
generators $J_{\mu \nu}$ acting through the usual untwisted
co-multiplication $\Delta_0$. Results from classical invariant
theory of orthogonal groups state that the subalgebra of $\cU$
invariants in $gr\cA$ is the polynomial algebra generated by the
image $(X^2)_0 \in gr\cA$ of the element
\[ X^2:=\sum_{\mu, \nu} \eta_{\mu \nu} x^\mu x^\nu \in \cA. \]
A simple calculation shows that $J_{\mu \nu}(X^2) =0$ for all $\mu$
and $\nu$, and this in turn leads to $J_{\mu \nu} (X^2)^k=0$ for all
$k$. Now let $\cA^0$ be the subset of $\cA$ consisting of elements
annihilated by all $J_{\mu \nu}$, that is,
\[
\cA^0= \{\phi\in\cA \mid J_{\mu \nu}(\phi)=0, \forall \mu, \nu\}.
\]
If $\phi\in\cA^0$ belongs to $\cA_i$ but not to $\cA_{i-1}$, then
its image in $gr\cA$ is a polynomial of degree $i$ in the variable
$(X^2)_0$. Then there exists some complex number $c$ such that
$\phi-c (X^2)^k\in \cA_{i-1}\cap \cA^0$. By induction on $i$ we can
show that $\cA^0$ in fact consists of polynomials in $X^2$.
Therefore, we have the following result:
\begin{quote}
the set $\cA^0$ of Lorentz invariant elements of $\cA$ consists of
polynomials in $X^2$ and thus forms a subalgebra of $\cA$.
\end{quote}

\begin{remark}
One may think that the result is intuitively clear, but in fact this
is far from the truth because the First Fundamental Theorem of
invariant theory breaks down in the present situation as the algebra
$\cA$ is noncommutative. Therefore, the result is quite interesting
mathematically from the point of view of invariant theory.
\end{remark}

\subsubsection{Algebra generated by matrix elements of a
representation} Let us now consider the subalgebra of $\cU^*$
generated by the matrix elements of the representation of $\cU$
associated to the module $V$. We shall denote this algebra by
$\cA(\fg)$.

Order the basis elements of $V$ as $x^0, x^1, x^2, x^3, \1$, and
denote $\1$ by $x^4$. Consider the matrix elements $t^a_b$ ($a, b=0,
1, \dots, 4$) of the representation of $\cU$ furnished by the module
$V$ relative to this basis. Here $t^a_b\in\cU^*$, such that for any
$u\in\cU$, $u x^a = \sum_{b=0}^4 t^a_b(u) x^b$. From $u x^4 = \epsilon(u)
x^4$, we obtain $t^4_4=\epsilon$, the co-unit of $\cU$. Also note
that $t^4_\mu=0$ for $\mu=0, 1, 2, 3$. A further property of the
matrix elements  is that if $\mu, \nu\le 3$,
\[
t_\mu^\nu (u P_\sigma) = t_\mu^\nu (P_\sigma u)=0, \quad  \forall
u\in\cU.
\]

Form the $5\times 5$ matrix $t=(t^b_a)$ where $a$ is the row index
and $b$ is the column index, and write $t(u)=(t^b_a(u))$ for any
$u\in\cU$. Then $t(u)t(u')=t(u u')$ for all $u, u'\in\cU$. Let
$C^{\mu \nu}:=\sum_{\sigma, \rho=1}^3 \eta_{\sigma \rho}
t^\sigma_\mu t^\rho_\nu$. Then $C^{\mu \nu}$ satisfies $C^{\mu
\nu}(u P_\mu)=0$ for all $u\in\cU$. Also $\sum_{\sigma, \rho=1}^3
\eta_{\sigma \rho} x^\sigma \otimes x^\rho$ is invariant under the
action of the Lorentz subalgebra, thus we conclude that
\begin{equation}\label{orthogonality}
\sum_{\sigma, \rho=1}^3 \eta_{\sigma \rho} t^\sigma_\mu t^\rho_\nu =
\eta_{\mu \nu}\epsilon.
\end{equation}
This is the familiar orthogonality relation satisfied by the matrix
elements of the natural representation of the orthogonal group.

It is easy to show that the opposite co-multiplication $\Delta'$ of
$\cU$ is related to $\Delta$ through
\[
\cF^{-2}\Delta = \Delta'\cF^{-2}
\]
where $\cF^{-2}$ satisfies all the defining properties of a
universal $R$-matrix. Thus it follows that
\begin{eqnarray}\label{RTT}
t_a^b t_c^d -  t_c^d  t_a^b  = - \delta_a^4 \delta_c^4
\left(i\theta^{b d}\epsilon - \sum_{\mu, \nu} i\theta^{\mu
\nu}t_\mu^b t_\nu^d\right),
\end{eqnarray}
where $\theta^{b d}=0$ if any of the indices is $4$. Now
(\ref{RTT}) is equivalent to the following
relations:
\begin{equation}\label{coordinates}
t^\nu_\mu t_c^d =  t_c^d t^\nu_\mu, \quad t_4^\mu t_4^\nu = t_4^\nu
t_4^\mu -i\theta^{\mu \nu}\epsilon - \sum_{\sigma, \rho=0}^3
i\theta^{\sigma\rho}t_\sigma^\mu t_\rho^\nu, \quad \mu, \nu\le 3.
\end{equation}
It follows from the first relation that the elements $t_\mu^\nu$
($\mu, \nu\le 3$) commute among themselves and also commute with all
the other matrix elements. In view of (\ref{orthogonality}), the
$t^\nu_\mu$ are nothing else but the matrix elements of the natural
module of the orthogonal group.

The second relation in (\ref{coordinates}) is reminiscent of the
relation (\ref{modulealgebra}). We may define $\omega^{\mu
\nu}:=\theta^{\mu \nu} \epsilon + \sum_{\sigma, \rho=0}^3
\theta^{\sigma\rho}t_\sigma^\mu t_\rho^\nu$. Then $\omega^{\mu
\nu}$ is skew symmetric in the indices $\mu$ and $\nu$ and commutes
with $t_4^\rho$ for all $\rho$. Also the components $\omega^{\mu
\nu}$ commute with one another. Denote $\zeta^\mu:=t_4^\mu$. Then
\begin{eqnarray}\label{omega}
\zeta^\mu \zeta^\nu - \zeta^\nu \zeta^\mu = i\omega^{\mu \nu}.
\end{eqnarray}

The elements $\zeta^\mu$ and $\omega^{\mu \nu}$ together generate a
subalgebra of $\cA(\fg)$. We may consider the commutative ring $R$
generated by all the components of $\omega^{\mu \nu}$, and consider
this subalgebra over $R$. Denote the $R$-algebra by $\cA_\omega$,
then again the relations (\ref{omega}) are the same relations as
those satisfied by the coordinate functions of the Moyal space but with
$\theta^{\mu \nu}$ replaced by $\omega^{\mu \nu}$.

One may be tempted to identify some completion of $\cA_\omega$
with the Moyal space, which however is not possible.
Note that $\cA_\omega$ is not stable under the action of the
Lorentz subalgebra corresponding to the left or right translations,
e.g., $\cL_{J_{\alpha \beta}}(\zeta^\mu\zeta^\nu)$ contains terms of
the form $t_\alpha^\mu\sum_\sigma \theta_\beta^\sigma t_\sigma^\nu$,
which do not belong to the subalgebra. This is not surprising since
the Lorentz generators do not generate a Hopf subalgebra of $\cU$.
It is also this fact which causes the induced module construction,
which works so well in the classical setting, to fail badly in the
context of the twisted Poincar\'e algebra.

\section{Noncommutative fields on Moyal space}\label{fields}

As we have already seen in Section \ref{noncommutative}, it is not
possible to generalise the induced module construction of Section
\ref{classical} to the noncommutative setting. This probably means
that there is no canonical definition of noncommutative fields in
relation to the representation theory of the twisted Poincar\'e
algebra.

However, we shall make a tentative proposal for the definition of
noncommutative classical fields on the Moyal space. It agrees with what
is assumed in the literature for scalar fields on the Moyal space. We hope
that this will provide a framework for studying twisted Poincar\'e
invariance of theories involving noncommutative fields with nonzero
spin.

A more systematic treatment of the problem addressed in this section
will require us to develop a theory of twisted Poincar\'e algebra
equivariant noncommutative vector bundles on the Moyal space. Even armed
with such a theory, one still needs to overcome, beside others,  the
difficulties discussed in Section \ref{noncommutative} in order to
have a satisfactory definition of noncommutative fields. Theses
matters are well beyond the scope of the present paper.

\subsection{Special type of commutative classical fields}\label{trivial}
Hereafter we denote by $\CR$ the space of complex valued smooth
functions on $\R^{1, 3}$. Denote by $x=(x^0, x^1, x^2, x^3)$ the
coordinate of $\R^{1, 3}$. Then regardless of what algebraic
structure we impose on $\CR$, we can always assign numerical values
to the $x^\mu$ to obtain numbers from elements of $\CR$. This is in
sharp contrast to the situation of Section \ref{modalg}.

We return to Section \ref{classical}, and consider the associated
vector bundle $\tilde G\times_L V \longrightarrow\tilde G/L$ in the
special case when
\begin{quote}
$V$ is a finite dimensional module for the twisted Poincar\'e
algebra $\cU$ with {\em trivial action} of all the generators
$P_\mu$.
\end{quote}
The bundle is trivial, thus its space of sections $\Gamma(V)$ is a
free module over $\left(C^\infty(\tilde G)\right)^{\cL_L}$. Note
that
\begin{eqnarray}\label{functions}
\CR= \left(C^\infty(\tilde G)\right)^{\cL_L},
\end{eqnarray}
and this is an identification of commutative associative algebras if
we equip $\CR$ with the usual commutative multiplication, which
shall be denoted by $\cdot$. Therefore, we have the $(\CR, \cdot)$-module isomorphism
\[ \Gamma(V) \cong \CR\otimes V. \]

In the special case under consideration, we can easily describe the
isomorphism. Now $C^\infty(\tilde G)$ contains a subalgebra which is
spanned by the matrix elements of the finite dimensional
representations of $\tilde G$ with trivial actions of all $P_\mu$.
Denote this algebra by $\cA(\fl)$. Then $\cA(\fl)$ in fact has the
structure of a commutative Hopf algebra.

Being a finite dimensional $\tilde G$-module, $V$ forms a right
$\cA(\fl)$ co-module. We denote the co-module map by
\[ \delta: V \longrightarrow V\otimes \cA(\fl), \]
and also use Sweedler's notation $\delta(v)=\sum_{(v)}
v_{(1)}\otimes v_{(2)}$ for any $v\in V$.  Then the isomorphism is
given by
\[
\psi: \CR\otimes V\longrightarrow \Gamma(V),
\quad a\otimes v \mapsto \sum_{(v)} v_{(2)} a \otimes v_{(1)}.
\]
It is a useful exercise to check that the image of $\psi$ is indeed
contained in $\Gamma(V)$, but we omit the details and refer to
\cite{GZ} for general ideas. Since $\CR$ is a $\tilde G$-module
under the action $\cL$, $\CR\otimes V$ admits a natural $\tilde G$
action via the usual co-product. One can easily show that $\psi$ is
a $\tilde G$-module map when $\Gamma(V)$ is regarded as a
$\tilde G$-module in the sense of (\ref{action}).

Elements of $\Gamma(V)$ are a special class of classical fields
determined by the inducing module $V$ of the spinor group, which is
in fact the restriction of a module for the entire Poincar\'e group
$\tilde G$.  The reason for us to consider this special case is that
this generalises to the noncommutative setting.

\subsection{Generalisation to noncommutative setting}\label{nc-fields}

Let us equip $\CR$ with the standard $\ast$-product defined for any
functions $f$ and $g$ by
\[
(f\ast g)(x) = \lim_{y\to x} \exp\left(\frac{i}{2}\sum_{\mu,
\nu}\theta^{\mu\nu} \frac{\partial}{\partial x_\mu}
\frac{\partial}{\partial y_\nu}\right)f(x) g(y), \quad x, y\in
\R^{1, 3}.
\]
Then $(\CR, \ast)$ is a noncommutative associative algebra. There is
the natural action of the twisted Poincar\'e algebra on $\CR$ given
by
\begin{eqnarray}
\begin{array}{r l}
&P_\mu(f)(x)  = -i \partial_\mu f(x), \\
&J_{\mu \nu}(f)(x) = -i x_\mu \partial_\nu f(x) + i x_\nu
\partial_\mu f(x).
\end{array}
\end{eqnarray}
By modifying this action we obtain another action $\varpi:
\cU\otimes \CR \longrightarrow \CR$ of the twisted Poincar\'e
algebra on $\CR$ given by
\[\varpi(u_1 u_2)(f) = S(u_2)( S(u_1)(f)) \]
for all $u_1, u_2\in\cU$ and $f\in\CR$.

As was first pointed out in \cite{CKNT} and very well known by now,
$\CR$ has the structure of a module algebra over the twisted
Poincar\'e algebra as a Hopf algebra with the twisted
co-multiplication $\Delta$. For any elements $f$ and $g$ of $\CR$,
and any $u\in \cU$,
\begin{eqnarray}
u(f\ast g) = \sum_{(u)} u_{(1)}(f)\ast u_{(2)}(g).
\end{eqnarray}
It also follows that $\CR$ has the structure of a module algebra
over $\cU$ under the action $\varpi$ with respect to the opposite
twisted co-multiplication $\Delta'$:
\begin{eqnarray}\label{varrho}
\varpi(u)(f\ast g) = \sum_{(u)} \varpi(u_{(2)})(f)\ast
\varpi(u_{(1)})(g).
\end{eqnarray}

A {\em noncommutative  scalar field} is an element $\phi$ of $\CR$
regarded as a $(\cU, \varpi)$ module, where $\phi$ vanishes rapidly
at infinity. This definition is in agreement with what implied in
the literature on noncommutative field theory, and reduces to the
usual definition of scalar fields in the commutative setting.

Two observations are important for the proposal of a definition of
noncommutative fields with nonzero spin. One is that the space
$\cA(\fl)$ of matrix elements of the finite dimensional
representations of $\cU$ with trivial $P_\mu$ actions for all $\mu$
forms a commutative subalgebra of the dual $\cU^*$ of the twisted
Poincar\'e algebra, and furthermore, $\cA(\fl)$ commutes with all
elements of $\cU^*$. Another observation is that there exists a
canonical {\em vector space} embedding $j: \CR\longrightarrow
C^\infty(\tilde G)$ given by equation (\ref{functions}) as a subset
of functions on the classical Poincar\'e group, since the algebraic
structure with the $\ast$-product is imposed afterward. Now
$\cA(\fl)\otimes \CR$ naturally has an associative algebra structure
with the multiplication, which we still denote by $\ast$, given by
\[
(a\otimes f)\ast (b\otimes g)= ab \otimes f\ast g
\]
for any $a\otimes f$ and $b\otimes g$ in $\cA(\fl)\otimes \CR$.
Consider the vector space embedding
\[
\begin{array}{l l l}
&i: \cA(\fl)\otimes \CR \longrightarrow C^\infty(\tilde G), \\
&i(a\otimes f)(\Lambda \exp(iP x)) = a(\Lambda) f(x), &\forall
\Lambda \exp(iP x)\in \tilde G,
\end{array}
\]
and denote
\[ \cA: = i(\cA(\fl)\otimes \CR). \]
We can introduce a noncommutative algebraic structure on $\cA$ by
setting
\[ i(a\otimes f)\ast i(b\otimes g) = i(a b\otimes f\ast g). \]
Then obviously $i$ is an algebra isomorphism between
$\cA(\fl)\otimes \CR$ and $\cA$, and we shall denote the resulting
algebra by $(\cA, \ast)$.

The twisted Poincar\'e algebra $\cU$ acts on $\cA(\fl)\otimes \CR$
\begin{equation}
\begin{array}{r l}
{\mathfrak R}: \cU\otimes \cA(\fl)\otimes \CR &\longrightarrow
\cA(\fl)\otimes \CR,\\ u\otimes a \otimes f \mapsto {\mathfrak
R}_u(a \otimes f)&=\sum_{(u)} \cR_{u_{(1)}}(a)\otimes
\varpi(u_{(2)})(f).
\end{array}
\end{equation}
This leads to a well defined  action on $\cA$
\begin{eqnarray}
\hat{\mathfrak R}: \cU\otimes \cA \longrightarrow \cA,
\end{eqnarray}
given for any $g=i(a\otimes f)$ with $a\otimes
f\in\cA(\fl)\otimes\CR$ by
\[
u\otimes i(a \otimes f) \mapsto \hat{\mathfrak R}_ui(a \otimes
f)=i\left(\sum_{(u)} \cR_{u_{(1)}}(a)\otimes
\varpi(u_{(2)})(f)\right).
\]
This turns $(\cA, \ast)$ into a module algebra for the twisted
Poincar\'e algebra.

\begin{remark}
The $\hat{\mathfrak R}$ action on $\cA$ can in fact be obtained by
differentiating the right translation by the Poincar\'e group.
\end{remark}

Any finite dimensional $\cU$-module $V$ with trivial actions of all
$P_\mu$ automatically has an $\cA(\fl)$ co-module structure, which
we still denote by
\[ \delta: V \longrightarrow V\otimes \cA(\fl), \quad v\mapsto \sum_{(v)}
v_{(1)}\otimes v_{(2)}.
\]
Define the map
\begin{eqnarray}
\psi_\theta: \CR\otimes V \longrightarrow  \cA \otimes V,
\end{eqnarray}
by $a\otimes v\mapsto \sum_{(v)}  i(v_{(2)}\otimes a)\otimes
v_{(1)}$, and set
\begin{eqnarray}\label{sections}
\Gamma_\theta(V) := \psi_\theta\left(\CR\otimes V\right),
\end{eqnarray}
where we emphasize again that \begin{quote} $V$ is assumed to be a
finite dimensional $\cU$-module with trivial actions for all $P_\mu$.
\end{quote} Then $\Gamma_\theta(V)$ forms a $\cU$-module with the
action defined for any $u\in\cU$ and $\zeta\in\Gamma(V)$ by
\[
u\cdot\zeta:=(\hat{\mathfrak R}_u\otimes \id_V)\zeta.
\]
Regard $\CR$ as a $\cU$-module with the action $\varrho$. Then
$\CR\otimes V$ has a natural $\cU$-module structure. It can be shown
that $\psi_\theta$ is $\cU$ linear.

For any element $\zeta=\sum_i g_i \otimes v_i$ of $\Gamma_\theta(V)$
and the special type of elements $\exp(iPx)\in \tilde G$,  we write
\[
\zeta(x):=\sum_i v_i g_i(\exp(iPx)).
\]
Then
\begin{eqnarray}\label{noncommutative-action}
(u\cdot\zeta)(x) = \sum  (\varpi(u_{(1)}) g_i)(x) u_{(2)}(v_i)  .
\end{eqnarray}

Note that if we rewrite the action (\ref{classical-action}) of the
Poincar\'e group on commutative classical fields in terms of the the
universal enveloping algebra of the Poincar\'e algebra, the
resulting formula will have the same form as
(\ref{noncommutative-action}).

Therefore, elements $\zeta$ of $\Gamma_\theta(V)$ may be
regarded as {\em noncommutative classical fields} determined by $V$.
[We have excluded internal degrees of freedom throughout the paper.]
For example, a noncommutative spinor field $\zeta$ is an element of
$\Gamma_\theta(V)$ if $V$ is the spinor module. However, we should
note that when all $\theta^{\mu \nu}=0$, this definition of fields
reduces to a special case of that in the commutative setting over
the usual Minkowski space.

\begin{remark}
The discussion after equation (\ref{tensor}) at the end of Section
\ref{classical} generalises to the noncommutative setting for the
$\Gamma_\theta(V)$ defined by (\ref{sections}).
\end{remark}

\subsection{The quantum case}
Finally we make a remark on the quantum case. After quantisation all
$g_i$ in a field $\zeta=\sum g_i\otimes v_i$ become operators (field
operators) acting on some Hilbert space. Denote the algebra of
field operators by $\cO$. Then every $g_i$ belongs to the algebra
$\cO\otimes\CR$ with the natural algebraic structure of the tensor
product of two algebras:
\[
(A\otimes f)(B\otimes h) = AB\otimes f\ast h, \quad \forall A,
B\in\cO, \ f, h\in\CR.
\]
The twisted Poincar\'e algebra is realised in terms of the field
operators $\iota: \cU\longrightarrow \cO$. In order for the action
of the twisted Poincar\'e algebra on $\cO$ to respect the algebraic
structure of the latter, one has to define the action of $\cU$ on a
quantum field $\zeta$ by
\begin{eqnarray}\label{dot}
(u\cdot \zeta)(x) := \sum   \iota(u_{(1)}) g_i(x) \iota(S(u_{(2)}))
\otimes v_i, \quad u\in\cU.
\end{eqnarray}
One can show that this indeed defines an action of $\cU$ on quantum
fields by noting that the right hand side involves the well-known
adjoint action of a Hopf algebra.

The transformation rule of the quantum field is then given by
\begin{eqnarray}
(u\cdot\zeta)(x) = \sum (\varpi(u_{(1)}) g_i)(x)\otimes u_{(2)}(v_i)
,
\end{eqnarray}
which is formally of the same form as (\ref{noncommutative-action}),
but with the left hand side given by (\ref{dot}) and the
$\varpi(u_{(1)})$ on the right side acting on the $\CR$ component of
$g_i$ only.

\section{Conclusion and outlook}\label{conclusion}

As we mentioned earlier, some researchers were clearly aware of the
necessity of formulating a precise transformation rule for
noncommutative fields under twisted Poincar\'e algebra. For example,
this was discussed at length by Fiore and Wess in \cite[Section 4]{FW}.
Lacking such a rule,  they suggested \cite{FW}
to replace it by a condition imposed on the Wightman functions.
 This condition formally looked the same as that in the
commutative case. Even assuming that one would eventually find any
justification for this, there is still the need of a
general rule to associate a field with a representation of the
Lorentz subalgebra of the twisted Poincar\'e algebra in order to
state the condition. So in this sense the transformation rule for
noncommutative fields under twisted Poincar\'e algebra can not be
entirely avoided. Our proposal for such a transformation rule in Section \ref{fields} is self
consistent, and should be the correct form. It hopefully provides the necessary framework for
studying twisted Poincar\'e invariance of noncommutative quantum
field theories on the Moyal space.

Another useful aspect of results reported here is that they point
out a way to look for possible noncommutative Minkowski spaces
suitable for developing quantum field theory with spacetime
symmetries described by Hopf algebras which are deformations of the
Poincar\'e algebra. In order for fields to naturally emerge within
such a framework, one might require the deformed Poincar\'e algebra to contain the
enveloping algebra of the Lorentz algebra or a deformation of it as
a Hopf subalgebra.

As an example, we consider the Poincar\'e algebra twisted by
\[ \cF_\tau=\exp(-i\tau J_{1 2}\otimes J_{3 4}).\]
[We could have anti-symmetrised the exponent as in \cite{LW} but have
not done so because we want to have simple expressions for the $\ast$-product.]
Now the universal enveloping algebra of the Lorentz
algebra is contained in this twisted Poincar\'e algebra as a Hopf
subalgebra. Thus there exists a noncommutative analogue $M_\tau$ of
the homogeneous space $\tilde G/L$ (the usual Minkowski space), and
noncommutative fields then naturally emerge as sections of
noncommutative homogeneous vector bundles on $M_\tau$. The
noncommutative Minkowski space $M_\tau$ is also easy to described.
Write $\cF^{-1}_\tau=\sum F_\alpha\otimes G_\alpha$. Define the
following noncommutative product $\ast_\tau$ on the space of
functions on $\R^{1, 3}$,
\[ (f\ast g)(x) = \sum (F_\alpha f)(x) (G_\alpha g)(x),  \]
and denote the resulting algebra by $(\CR, \ast_\tau)$. Then $(\CR,
\ast_\tau)$ is the algebra of functions on the noncommutative
Minkowski space. Denote by $X_\mu$, $\mu=1, 2, 3, 4$, the coordinate
functions, that is
\[ X_\mu(x) = x_\mu, \quad x\in \R^{1, 3}. \]
Then $(X_\alpha\ast_\tau X_\beta)(x) = x_\alpha x_\beta$, if
$\alpha\ge \beta$, or $\alpha, \beta\in \{1, 2\}$, or $\alpha,
\beta\in \{3, 4\}$; and
\begin{eqnarray}
\begin{array}{r l l}
(X_1\ast_\tau X_3)(x) &=&  x_1 x_3\cos\tau - i  x_2 x_4\sin\tau, \\
(X_1\ast_\tau X_4)(x) &=&  x_1 x_4\cos\tau + i  x_2 x_3\sin\tau, \\
(X_2\ast_\tau X_3)(x) &=&  x_2 x_3\cos\tau + i  x_1 x_4\sin\tau, \\
(X_2\ast_\tau X_4)(x) &=&  x_2 x_4\cos\tau - i  x_1 x_3\sin\tau.
\end{array}
\end{eqnarray}
It will be interesting to construct quantum field
theoretical models on such a noncommutative Minkowski space,
which are invariant with respect to the twisted Poincar\'e algebra.

Other possible examples are the quantum Poincar\'e algebras
constructed in \cite{OSWZ,
CD} in the context of the complexified
conformal algebra $so(6, \C) = sl(4, \C)\subset gl(4, \C)$. These
quantum Poincar\'e algebras are quantised parabolic subalgebras of
the enveloping algebra of $gl(4, \C)$, and contain the quantum group
$\U_q(sl_2)\otimes \U_q(sl_2)$ as a Hopf subalgebra, which is the
quantised version of the enveloping algebra of the complexified
Lorentz algebra. In these cases, there exist natural quantum
homogeneous spaces which play the role of the Minkowski space. The
noncommutativity of the quantum Minkowski spaces is now much more
severe than that of the standard Moyal space or the
previous example. Nevertheless by using appropriate analogues of the
quantum Haar measure \cite{GZ} one may be able to construct
quantum Poincar\'e invariant field theory on such quantum Minkowski
spaces.

There remains the possibility that the Seiberg-Witten
map \cite{SWit} allows for a realisation of spacetime symmetry of
the twisted Poincar\'e type.  We also mention that quantum group symmetries
manifest themselves in conformal field theory as well \cite{ACS}, but
in a manner different
from spacetime symmetries. It will be interesting to understand
such quantum group symmetries from the point of view of
noncommutative geometry.

\acknowledgments We are indebted to Jos\'e Gracia-Bond\'ia, Jerzy
Lukierski  and Peter Pre\v{s}najder for helpful comments. P. P.
Kulish, R. B. Zhang and X. Zhang wish to thank the Department of
Physical Sciences, University of Helsinki, and the Helsinki
Institute of Physics for the hospitality extended to them during
their visits in September-October 2007 when this work was completed.
Partial financial support from the Australian Research Council,
National Science Foundation of China (grant 10421001), NKBRPC
(2006CB805905), the Chinese Academy of Sciences and the Academy of
Finland is also gratefully acknowledged.

\end{document}